\documentclass[%
 reprint,
 amsmath,amssymb,
 aps,
]{revtex4-2}

\usepackage{graphicx}
\usepackage{dcolumn}
\usepackage{bm}
\usepackage{csquotes}
\usepackage{soul,xcolor}
\sethlcolor{yellow}

\begin{document}

\preprint{APS/123-QED}

\title{Signatures of two ferromagnetic states and goniopolarity in LaCrGe$_3$ in the Hall effect}
\author{
Modhumita Sariket$^1,$
Najrul Islam$^1,$
Ayan Jana$^1$, Manoranjan Kumar$^1,$
Saquib Shamim$^1$}%
\author{Nitesh Kumar$^{1,}$}%
 \email{Contact author: nitesh.kumar@bose.res.in}
\affiliation{%
 $^1$Department of Condensed Matter and Materials Physics,\\S. N. Bose National Centre for Basic Sciences, Salt Lake City, Kolkata-700106, India
}

 \begin{abstract}
LaCrGe$_3$ has become a playground to understand quantum critical phenomena in ferromagnetic (FM) materials. It has also garnered attention due to its peculiar two FM phases. Here, we demonstrate the presence of these phases using the Hall effect. Continuous temperature-dependent Hall resistivity measurements at fixed magnetic fields clearly demonstrate the presence of these phases, regardless of the direction of the applied magnetic field. The remanent Hall resistivity and Hall coefficient undergo a maximum and a minimum, respectively, at the boundary between the two phases. We observe a significantly large anomalous Hall conductivity of 1160 $\Omega^{-1}$cm$^{-1}$ at 2 K when the magnetic field is applied along the magnetic easy axis, which is dominated by intrinsic effects, at least in the low-temperature FM phase. In the paramagnetic (PM) phase, hexagonal LaCrGe$_3$ exhibits opposite charge carrier polarities along different crystallographic directions, attributed to the anisotropic Fermi surface geometry, a phenomenon known as \enquote{goniopolarity}. The coexistence of goniopolar transport and unconventional magnetic phases may lead this material as a promising candidate for future electronic devices.

\end{abstract}
\maketitle
\section{\label{sec:level1}Introduction:}
Itinerant ferromagnets are a fertile platform for exploring unconventional magnetic states where spin order emerges from the interplay between electronic correlations, crystal symmetry, and Fermi-surface topology \cite{buschow2003itinerant}. These systems are often studied to search for novel states in the vicinity of quantum critical point \cite{chubukov2004instability,huang2015anomalous}. An itinerant ferromagnet LaCrGe$_3$ displays a rich temperature-pressure-magnetic field phase diagram that includes two FM phases, a tri-critical wing structure, and two antiferromagnetic (AFM) phases adjoining the PM region \cite{lin2013suppression,taufour2018ferromagnetic,krenkel2024enhancement,bosch2021magnetic,kaluarachchi2017tricritical,belitz2017quantum,taufour2016ferromagnetic}. This indicates that the intricate magnetic behavior originates from the itinerant Cr 3\textit{d} spin structure, coupled with unusual domain wall dynamics. Such domain wall dynamics can give rise to multiple ordered phases by tuning temperature alone. 

 \vspace{2 mm}LaCrGe$_3$ is a collinear ferromagnet with a Curie temperature $T_\mathrm{C}$=85 K, where the existence of two distinct FM phases has been proposed. The two FM (FM$_1$ and FM$_2$) phases were first assigned based on a broad maximum in the temperature derivative of the in-plane resistivity \cite{kaluarachchi2017tricritical}. A similar behavior, indicative of two FM states, has also been reported at ambient pressure in UGe$_2$ based on resistivity measurements \cite{hardy2009two}. Additional experimental evidence for the presence of two FM phases in LaCrGe$_3$ has been provided by magnetization, resistivity, and muon spin relaxation studies \cite{rana2021magnetic,gati2021formation,xu2023unusual}. However, techniques such as electron spin resonance, neutron diffraction, and thermodynamic measurements have not revealed any clear signatures of two distinct FM phases \cite{sichelschmidt2021electron,das2014heavy}. Beyond its ordered states, LaCrGe$_3$ also exhibits intriguing properties in the PM phase where nuclear magnetic resonance (NMR) studies reveal isotropic FM-like spin fluctuations, suggesting a higher degree of electron localization compared to typical itinerant Cr 3\textit{d} systems \cite{rana2019magnetic}.

 \vspace{2 mm}Despite several reports on the quantum critical behaviour and attempts to probe magnetism in LaCrGe$_3$, Hall effect studies have remained largely unexplored. The study of Hall effect in a ferromagnet in the ordered and paramagnetic states can reveal intricate details of magnetism and electronic structure. Here, we employ magneto-transport measurements in particular the Hall effect to substantiate the existence of two FM phases in LaCrGe$_3$ since it is highly sensitive to changes in the electronic band structure and spin-dependent scattering associated with magnetic phase transitions. Our detailed treatment of Hall effect unravels the existence of two FM phases in terms of sharp anomalies in the continuous temperature dependent Hall resistivity at fixed magnetic fields, a maximum in remanent Hall resistivity and a minimum in Hall coefficient at the boundary between two FM phases. Interestingly, we observe a so called goniopolar effect in terms of direction-dependent conduction polarity in LaCrGe$_3$ in the PM state by Seebeck effect combined with Hall effect studies. The density functional theory yields high anisotropy in the Fermi surface which supports our observation of goniopolarity.  
\section{EXPERIMENTAL METHODS}
Single crystals of LaCrGe$_3$ were synthesized by the self-flux method. First, a polycrystalline ingot of LaCrGe$_3$ was prepared by arc melting a stoichiometric amount of La (AlfaAesar, 99.9 \%), Cr  (Thermo Fisher, 99.99 \%), and Ge (AlfaAesar, 99.9999 \%). An excess of Ge was added as flux to the powdered LaCrGe$_3$, with the total molar ratio maintained at 13:13:74 (La:Cr:Ge) [8]. The content was placed in an alumina crucible within a quartz tube, which was sealed under vacuum.The mixture was heated to 1100 °C at a rate of 200 °C/h and kept at this temperature for 5 h, followed by cooling to 825 °C in 65 h. The mixture was heated to 1100 \textdegree C at a rate of 200 \textdegree C/h and kept at this temperature for 5 h, followed by cooling to 825 \textdegree C in 65 h. At this temperature, excess flux was removed by centrifugation. Millimeter-sized shiny hexagonal rod-shaped single crystals were obtained. Energy-dispersive X-ray spectroscopy (EDX) data were collected to confirm the chemical composition using a field emission 
\begin{figure*}[t!]
    \centering
  \includegraphics[width=1\textwidth]{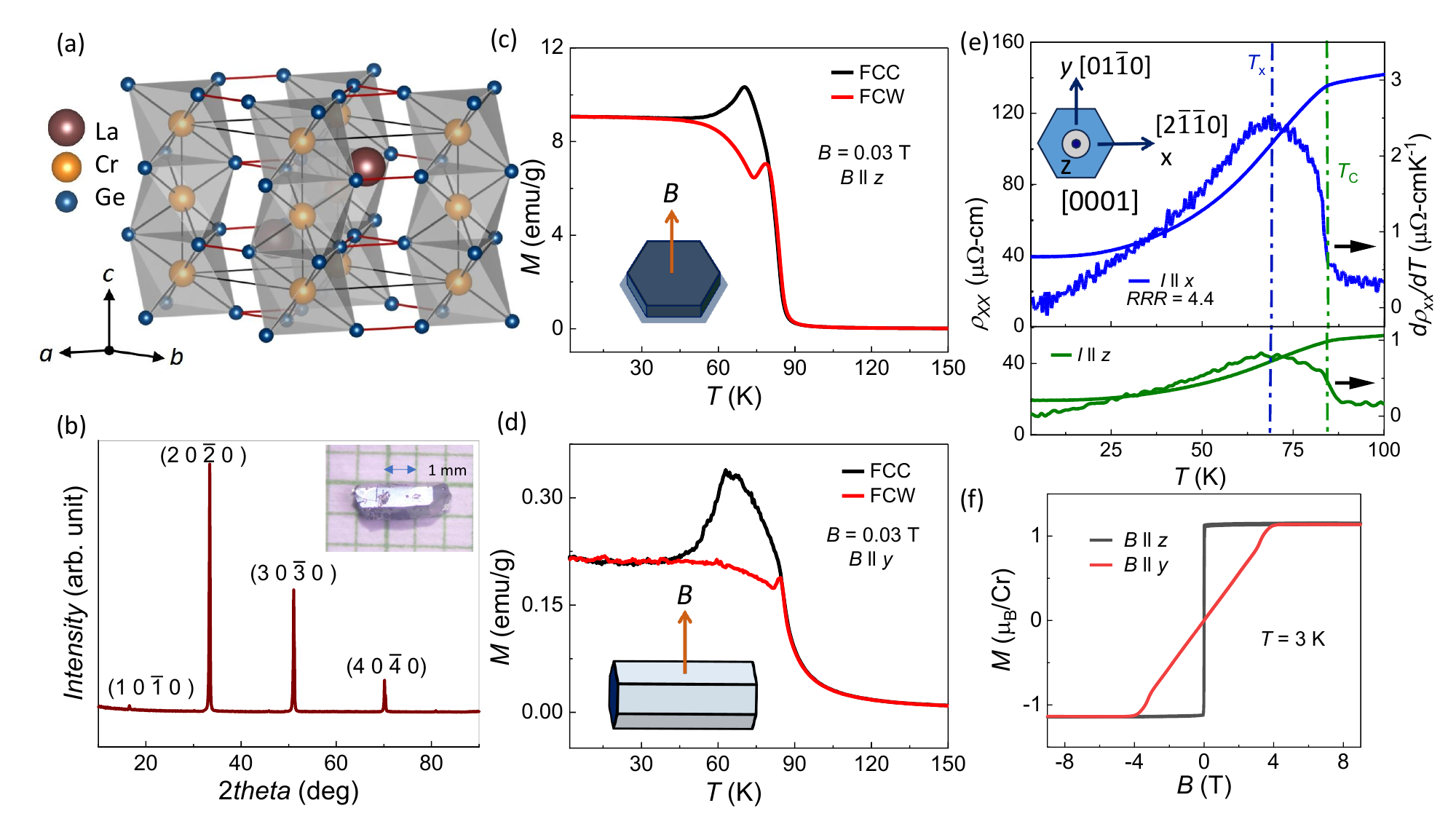}
    \caption{(a) Crystal structure of LaCrGe$_3$. (b) XRD pattern obtained by exposing the \textit{ac}-plane of the crystal (see inset) to the x-ray beam. (c)-(d) Magnetization as a function of temperature measured with field cooled cooling (FCC) and field cooled warming (FCW) protocols with magnetic field applied along (c) \textit{z} axis and (d) \textit{y} axis. (e) Temperature dependence of the resistivity and its temperature derivative with the current applied along the \textit{x}- and \textit{z}-axes, shown by the blue and green curves, respectively. The inset defines the \textit{x}, \textit{y} and \textit{z}-axes with respect to the crystallographic directions. Green and blue vertical lines represent  $T_\mathrm{C}$ and $T_\mathrm{x}$. (f) \textit{M}-\textit{B} isotherms at 3 K for \textit{$B || z$} and \textit{$B || y$} axes.}
    \label{fig:your-label}
\end{figure*}
scanning electron microscope (Quanta 250 FEG) equipped with a silicon drift detector (SDD) at 25 kV accelerating voltage. X-ray diffraction (XRD) of the powdered single crystals was performed at room temperature using a Rigaku SmartLab diffractometer at 9 kW with Cu K$\alpha$ radiation. XRD pattern was refined using the FullProf software. Magnetic and electrical transport measurements were performed using the VSM (vibrating sample magnetometer) option and the ETO (electrical transport option), respectively, in Dynacool, physical property measurement system by Quantum Design, USA. Temperature-dependent magnetic measurements were performed under zero field cooling (ZFC), field cooled cooling (FCC), and field cooled warming (FCW) conditions. Magnetization loops were recorded over four quadrants, sweeping the magnetic field (\textit{B}) from 9 T to -9 T. The standard four-probe method was followed to measure the longitudinal and Hall resistivity on rectangular bar-shaped crystals. Hall resistivity and longitudinal resistivity were anti-symmetrized and symmetrized, respectively, to remove the effect of probe misalignment.

 \vspace{2 mm}The Seebeck coefficients were measured using a dipstick setup housed in a liquid nitrogen cryostat. A controlled temperature gradient was applied across rectangular bar-shaped samples, using a Keithley 2400 source measure unit. The local temperatures at the ends of the sample were measured using PT100 sensors. The resulting thermovoltage was recorded using a high-precision nanovoltmeter (Keithley 2182A).
 
\vspace{2 mm}The electronic band structure and Fermi surface of LaCrGe$_3$ were further investigated within the framework of the density functional theory (DFT) using the Vienna ab initio simulation package (VASP) \cite{PhysRevB.54.11169}. Two types of exchange-correlation functions were employed: local density approximation (LDA) \cite{PhysRev.140.A1133} and the generalized gradient approximation (GGA) form in the Perdew-Burke-Ernzerhof (PBE) parameterization, in order to access the effect of exchange correlation on the electronic properties \cite{PhysRevB.45.13244,PhysRevLett.77.3865}.
For electronic band structure calculations, a $12 \times12 \times9$ $\Gamma$-centered Monkhorst-Pack  \textit{k}-point mesh was used \cite{PhysRevB.13.5188}. The plane-wave cutoff energy was set to 400 eV, and ionic relaxation was carried out until the forces were below 0.002 eV/\text{\AA}, with an electronic convergence criterion of $1 \times 10^{-8}$ eV. Ionic relaxation was performed using the conjugate gradient method \cite{hestenes1952cg}. The electronic structure calculations were done both in presence and absence of the spin-orbit coupling (SOC). For the Fermi surface calculations, a denser $15 \times 15 \times 12 $ \textit{k}-point grid was employed and these were calculated and visualized using the FermiSurfer software package \cite{fermisurfer2019}.
\section{RESULTS AND DISCUSSION}
\subsection{\label{sec:level2} Crystal Synthesis and Characterization}
 LaCrGe$_3$ adopts a hexagonal BaNiO$_3$-type crystal structure, which belongs to the space group \textit{P}6$_3$/\textit{mmc} \cite{cadogan2013neutron}. Fig. 1(a) depicts the crystal structure of LaCrGe$_3$, where Cr atoms occupy the centers of face-sharing CrGe$_6$ octahedra \cite{lin2013suppression,bie2007structures}. These octahedra are arranged along the \textit{c}-axis, forming one-dimensional chains of Cr atoms, with a nearest-neighbor Cr-Cr distance of approximately 2.88 \text{\AA}. In contrast, the shortest Cr-Cr distance in the \textit{ab}-plane is quite large ($\sim$6.2 \text{\AA}), resulting in a strongly anisotropic crystal structure. Each Cr atom is coordinated by Ge atoms with Cr-Ge bond lengths of around 2.52 \text{\AA}. Individual chains interact through Ge forming triangular clusters with Ge-Ge distances of about $\sim$ 2.6 \text{\AA}, resulting in a breathing kagome lattice within the basal plane. La atoms occupy the space between CrGe$_3$ columns to make the overall three-dimensional structure of LaCrGe$_3$. The relaxed structure obtained from DFT calculations agrees well with the experimental bond lengths. Fig. 1(b) shows the XRD pattern collected on a hexagonal rod-shaped crystal (inset of Fig. 1(b)) by keeping the rectangular face parallel to the sample holder. The presence of sharp peaks with only $(h\ 0\ \bar{h}\ 0)$ indices indicates that the exposed rectangular face is the \textit{ac}-plane. The quality as well as the orientation of the single crystals were further confirmed by Laue diffraction as shown in SI (Supplementary Information \cite{supplementary}). The Rietveld refinement of the powdered single crystals is consistent with the \textit{P}6$_3$/\textit{mmc} space group \cite{supplementary}.
 
 \vspace{2 mm}As described earlier, LaCrGe$_3$ is an unusual ferromagnet that not only undergoes a conventional PM to FM phase transition at $T_\mathrm{C}$=85 K, but also exhibits an additional FM transition ($T_\mathrm{x}$) at a lower temperature, indicative of a complex magnetic ordering \cite{xu2023unusual,kaluarachchi2017tricritical,ullah2023magnetic}. Figs. 1(c) and 1(d) show the temperature-dependent magnetization measured under FCC and FCW conditions, with an applied \textit{B} of 0.03 T along the \textit{z} and \textit{y} axes, respectively.
  \begin{figure}[t]
    \centering \includegraphics[width=0.5\textwidth]{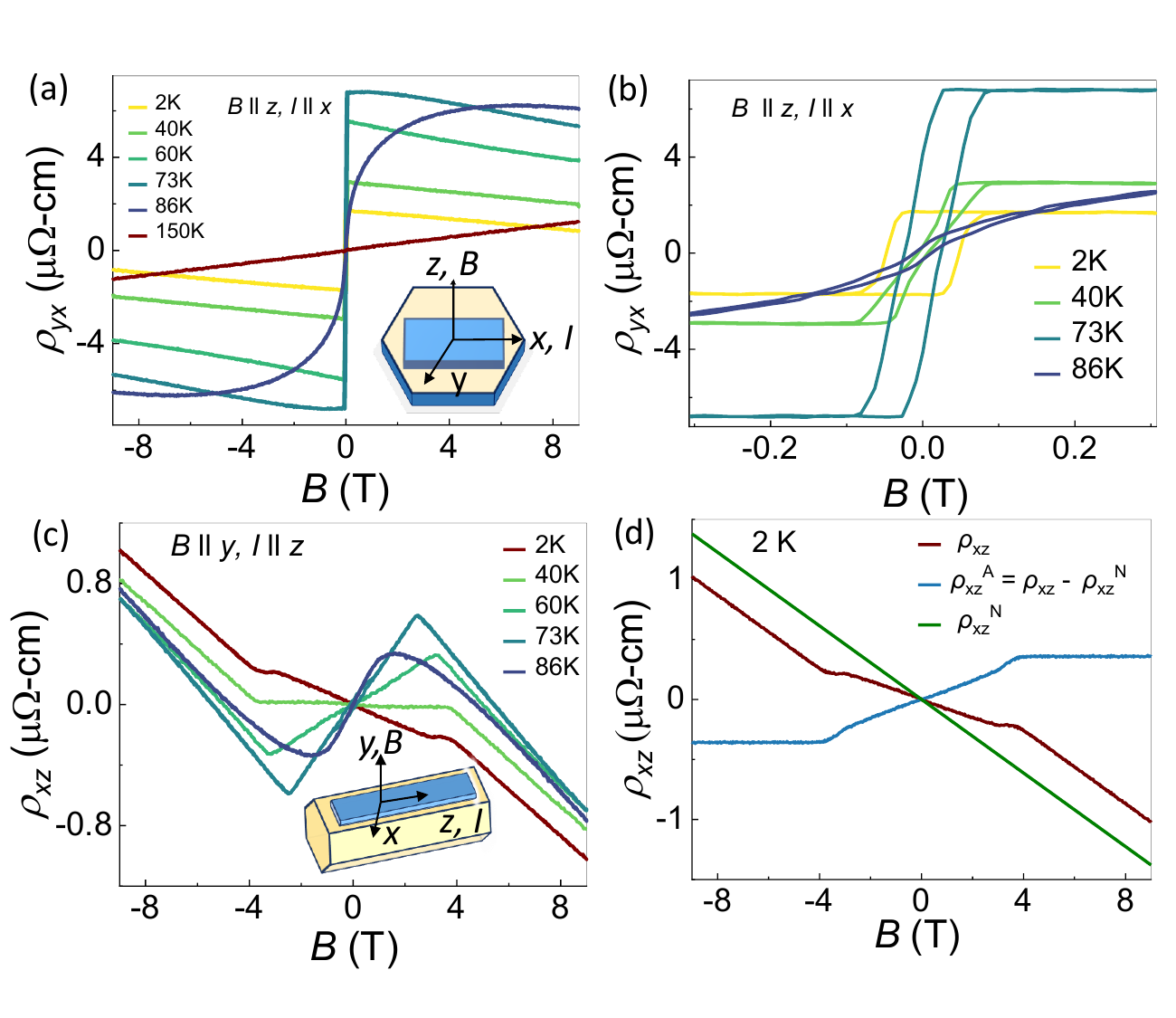}
   \caption{(a) Field-dependent Hall resistivity, $\rho_{yx}$, at various temperatures, measured while sweeping $B$ from -$9$ T to $9$ T. The inset illustrates the magnetic field and current directions (\textit{$B || z$}, \textit{$I || x$}). (b) Magnified view of the low-field region of $\rho_{yx}$ for \textit{$B || z$}, \textit{$I || x$} at selected temperatures. (c) Field-dependent $\rho_{xz}$ isotherms at different temperatures, with the inset showing the corresponding field and current directions (\textit{$B||y$}, \textit{$I || z$}). (d) Normal ($\rho_{xz}^N$) and anomalous ($\rho_{xz}^A$) components, along with the total Hall resistivity ($\rho_{xz}$), at $T$=2 K for \textit{$B || y$}, \textit{$I || z$}.}
    \label{fig:your-label}
\end{figure}
 As the temperature is decreased there is a steep increase in the magnetization due to the ferromagnetic ordering at $T_\mathrm{C}$ followed by a distinct kink near 70 K in the FCC curve. The FCW curve, however, differs from the FCC curve and features an anomaly characterized by a minimum at 75 K. The magnetization curves show similar features for \textit{$B || y$} as shown in Fig. 1(d). These features have been attributed to the presence of two FM transitions corresponding to two exchange constants \cite{ullah2023magnetic}. The decrease in magnetization in FCC below 70 K is unusual for a ferromagnet which is observed for both \textit{$B || y$} and \textit{$B || z$} directions.  In the absence of a structural phase transition, the hysteresis between ZFC and FCW curves can originate from spin reorientation or magnetic domain formation. In the case of LaCrGe$_3$, both magnetic measurements and neutron powder diffraction indicate that this behavior is better explained  by the domain wall pinning-depinning mechanism \cite{ullah2023magnetic,bosch2021magnetic}. In addition, no anomaly associated with the anisotropy constant with respect to temperature is observed in the magnetically ordered state \cite{ullah2023magnetic,supplementary,aharoni1998demagnetizing}. In this scenario, if multiple domain walls get pinned over a certain temperature range just below $T_\mathrm{C}$, a hysteresis between ZFC and FCW can be explained. Similar magnetization behavior due to domain wall pinning effect has also been observed in Co$_3$Sn$_2$S$_2$ and Fe$_3$GaTe$_2$ \cite{shen2022anomalous,birch2022history}. Temperature-dependent resistivity curves show a metallic nature with a residual resistivity ratio (\textit{RRR} = $\rho_\mathrm{300K}/\rho_\mathrm{2K}$) value of 4.4 with a kink feature near the FM ordering temperature (Fig. 1(e)). Only in the temperature derivative of the in-plane resistivity (\textit{d{$\rho$}$_{ab}$/dT}), a broad maximum is seen near $T_\mathrm{x}$ for both the current (\textit{I}) directions parallel to the \textit{x} and \textit{z} axes,  as previously reported \cite{kaluarachchi2017tricritical}. In Fig. 1(f), the magnetic moment saturates at a field of 0.07 T at 3 K when the field is applied parallel to the \textit{z}-axis, indicating that this is the direction of easy magnetization. In contrast, for the hard direction (\textit{$B||y$}-axis), an anisotropy field of about 4 T is required to achieve the saturation magnetic moment 1.25 $\mu_B$/Cr. This relatively small ordered magnetic moment compared to the effective moment in the paramagnetic (PM) state above $T_\mathrm{C}$, $\mu_{eff}$ = 2.4 $\mu_\mathrm{B}$/Cr, indicates significant delocalization of the Cr 3\textit{d} electrons \cite{takahashi1986origin}. However, NMR studies reveal signatures of Cr 3\textit{d} electron localization \cite{rana2019magnetic}, indicating a competition between localized, correlated \textit{d} orbitals and delocalized \textit{p} orbitals due to hybridization. This competition gives rise to different complex magnetic states under applied pressure \cite{wysokinski2019mechanism}. 
\subsection{\label{sec:level2} Magnetotransport Study}
 To gain insight into the electrical properties of LaCrGe$_3$, we performed transverse resistivity (${\rho_{ij}}$) measurements. For a typical ferromagnet, ${\rho_{ij}}$ consists of contributions from the normal Hall resistivity ($\rho_{ij}^{N}$), which arises from the magnetic Lorentz force on moving charges, and the anomalous Hall resistivity ($\rho_{ij}^{A}$) which is proportional to the sample magnetization (\textit{M}) and could consist both intrinsic and extrinsic mechanisms. These contributions can be expressed through the widely used empirical formula: ${\rho_{ij} = \rho_{ij}^{N} + \rho_{ij}^{A} = R_0B + \mu_0R_SM}$, where $R_0$ and $R_S$ are the ordinary Hall coefficient and the anomalous Hall coefficient, respectively, and $\mu_0$ is the vacuum permeability \cite{nagaosa2010anomalous}. $\rho_{yx}$ as a function of \textit{B} applied along the \textit{z}-axis (\textit{$I || x$}), is shown in Fig. 2(a). Below $T_\mathrm{C}$, we find the signature of anomalous Hall effect (AHE) in terms of steep increase of Hall resistivity at low magnetic field followed by a weak linear \textit{B} dependence. While the low field steep rise of Hall resistivity provides the value of anomalous Hall resistivity which has a linear magnetization dependence according to the above empirical formula, the slope in the high field region gives ordinary Hall coefficient. Above 100 K, i.e., above $T_\mathrm{C}$, $\rho_{yx}$ displays a linear field dependence in the whole field range without any slope change, characteristic of only ordinary Hall behavior. Fig. 2(b) shows the enlarged view of the data presented in Fig. 2(a) at small \textit{B} for some selected temperatures. At 2 K, the coercive field is 0.048 T, with a remanent Hall resistivity of 1.7 $\mu\Omega$-cm. Although the exact loop shape has a weak crystal dependence, the variation of the coercive field for a fixed crystal shows a peculiar temperature dependence, which we will discuss in the next section. Fig. 2(c) shows the Hall resistivity ($\rho_{xz}$) for \textit{$B || y$}-axis and \textit{$I || z$}-axis at different temperatures where $\rho_{xz}$ increases linearly and then abruptly changes the slope near 4 T. The first clear observation is that the magnitude of the slopes in the anomalous Hall and normal Hall regions are comparable. At first glance, anomalous Hall resistivity (AHR) seems to change sign from negative to positive as the temperature increases. However, when the contribution of the normal Hall effect is removed from the total Hall resistivity (see Fig. 2(d)), we recover positive AHR for all temperatures in the ordered state. The nature of the coercive field follows a similar trend with temperature as observed in the case of \textit{$B || z$}-axis as shown \cite{supplementary}.

 \begin{figure}[h]
    \centering
    \includegraphics[width=0.5\textwidth]{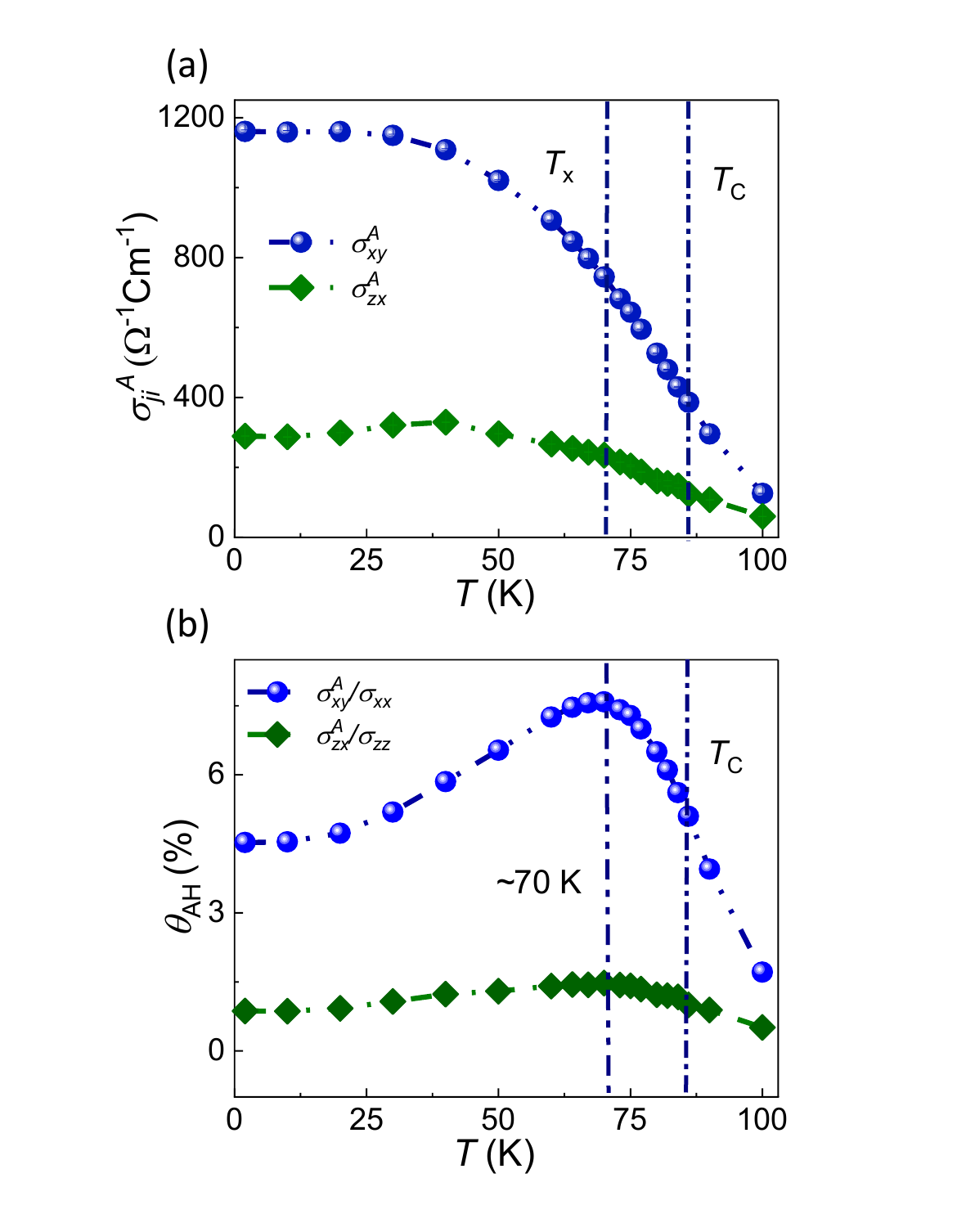}
    \caption{(a) Anomalous Hall conductivity ($\sigma_{ij}^A$) as a function of temperature (\textit{T}) for the field direction parallel to the \textit{z}-axis and \textit{y}-axis where two transition temperatures $T_\mathrm{C}$ and $T_\mathrm{x}$ are indicated. (b) depicts anomalous Hall angle ($\theta_\mathrm{AH}$) vs. temperature for both the directions. }
    \label{fig:your-label}
\end{figure}
 \begin{figure*}[t!]
    \centering
    \includegraphics[width=1\textwidth]{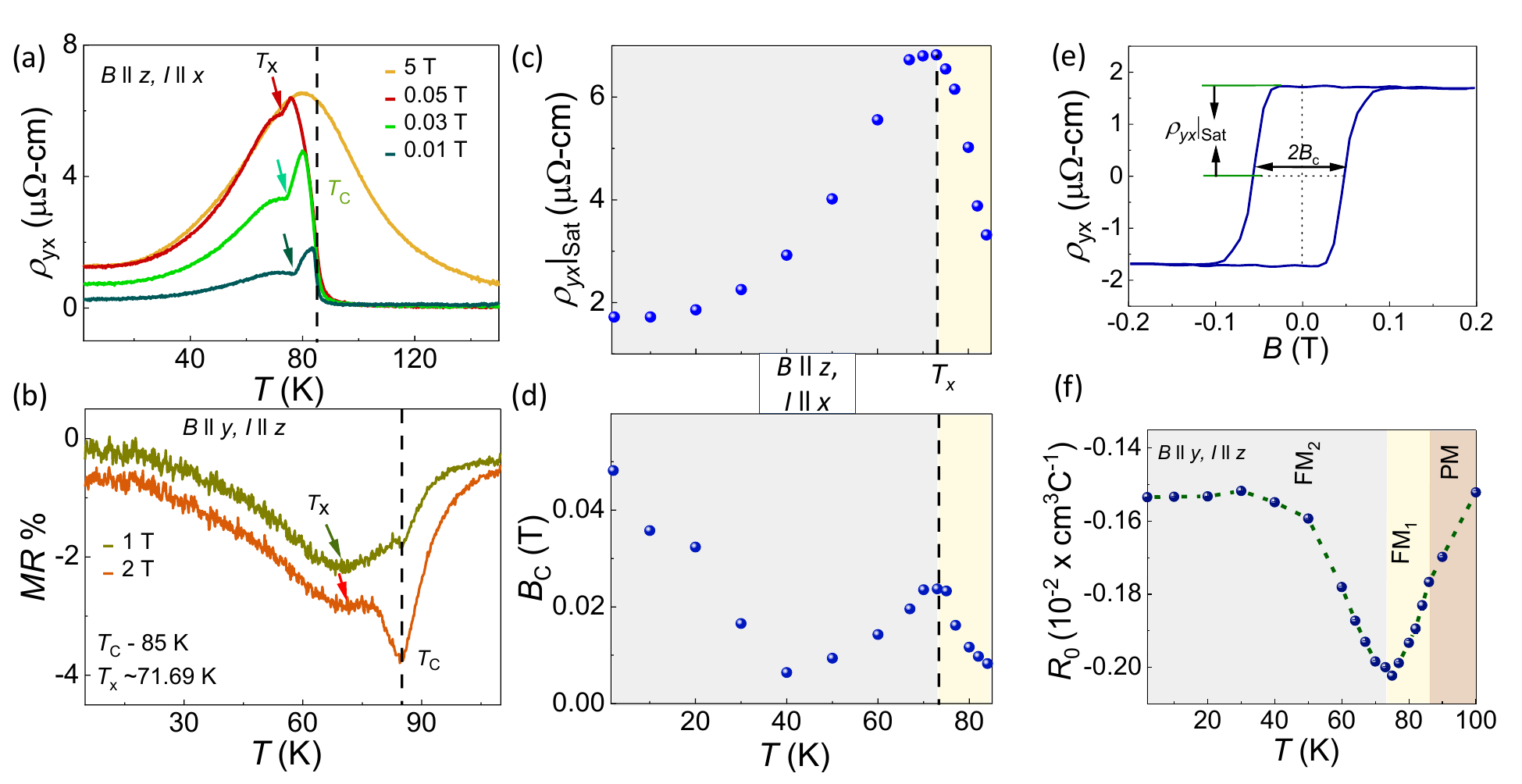}
    \caption{(a) Hall resistivity ($\rho_{yx}$) as a function of temperature (\textit{T}) during the heating time for the applied magnetic field 0.01 T, 0.03T, 0.05T and 5 T along the \textit{z}-axis where $T_\mathrm{C}$ is PM-FM$_1$ and $T_\mathrm{x}$ is FM$_1$-FM$_2$ transitions. (b) Magneto-resistance ($\text{\textit{MR}}\%$) with changing magnetic field during heating time for the applied magnetic field 1 T and 2 T the \textit{y}-axis where  $T_\mathrm{C}$ is indicated by the dashed line and $T_\mathrm{x}$ is indicated by the arrows. (c) and (d) $\rho_{yx}|_{Sat}$ and $B_\mathrm{c}$ are plotted against temperature for the applied field along the \textit{z}-axis where $T_\mathrm{x}$ is shown by a dashed line. (e) For a hysteresis loop of $\rho_{yx}$ verses \textit{B},  $\rho_{yx}|_{Sat}$ and $B_\mathrm{c}$ are indicated. (f) Ordinary Hall coefficient is plotted against temperature for the applied field along the \textit{$B || y$}-axis where PM, FM$_1$ and FM$_2$ regions are separated by different colors.}
    \label{fig:your-label}
\end{figure*}
 
For both $B || z$ and $B || y$-axis, $\rho_{ij}$ shows a negative slope in the high field regions below $T_\mathrm{C}$, indicating the electron dominant normal electrical transport. We calculate the Hall conductivity $\sigma_{ji}$ using the tensor relation, $\sigma_{ji} = \rho_{ij} /(\rho_{ij}^2 + \rho_{ii}*\rho_{jj})$. From the Hall conductivity isotherms ($\sigma_{ji}$-$B$ data), the anomalous Hall conductivity, $\sigma_{ji}^A$, was extracted as the \textit{y}-intercept of the high-field linear fit. These values are plotted as a function of temperature for both \textit{$B || y$} and \textit{$B || z$} in Fig. 3(a).  At the lowest measurement temperature of 2 K, we observe $\sigma_{zx}^A$ = 289 $\Omega^{-1}$cm$^{-1}$ (\textit{$B || y$}) and $\sigma_{xy}^A$= 1160 $\Omega^{-1}$cm$^{-1}$ (\textit{$B || z$}) which are higher than the values previously reported for the \textit{ab}-plane \cite{li2024giant}. With increasing temperature, $\sigma_{ji}^A$ remains nearly constant up to 50 K, followed by a gradual decrease near $T_\mathrm{C}$. The temperature-independent behavior of $\sigma_{ji}^A$ for a substantial range at low temperature for both directions indicates that the anomalous Hall conductivity (AHC) is dominated by the intrinsic contribution which can originate from the Berry curvature mechanism associated with several band crossings observed close to the Fermi energy (\textit{E}$_{\text{F}}$). To elucidate the elemental contributions to the electronic band structure, we have computed the projected band structure, together with the different dominant orbital contributions of La, Cr and Ge as shown in \cite{supplementary}. Notably, the dominant contribution near the \textit{E}$_{\text{F}}$ arises from Cr-derived states, while the hybridization is strongest between Cr and Ge in the vicinity of the \textit{E}$_{\text{F}}$. An important observation here is a significantly larger value of $\sigma_{xy}$ compared $\sigma_{zx}$ despite the larger $\rho_{xx}$ compared to $\rho_{zz}$ and equal saturation magnetization values. This signifies that anomalous Hall conductivity broadly arises from intrinsic mechanism and does not depend on scattering events which is the main reason for extrinsic skew scattering mechanism. A detailed scaling analysis of AHE is discussed in \cite{supplementary}. The anomalous Hall angle ($\theta_\mathrm{AH}$) defined as $\sigma_{ji}^A/\sigma_{jj}$ ($\%$) quantifies the relative strength of the anomalous Hall current compared to the longitudinal current \cite{nagaosa2010anomalous}. We observe a significantly large $\theta_\mathrm{AH}$, reaching value as large as 7.5 $\%$ ($\sigma_{xy}^A/\sigma_{xx}$) at 70 K for $B||z$-axis. Whereas, for $B||y$-axis, the value of $\theta_\mathrm{AH}$ is relatively small ($\sigma_{zx}^A/\sigma_{zz} =$ 2.1 $\%$ at 70 K). For both the $B||z$ and $B||y$ axes, $\theta_\mathrm{AH}$ peaks at $T_\mathrm{x}$ ($\sim$70 K) and decreases thereafter. This anomaly may correspond to a transition between two FM phases that will be discussed in the next section.

\subsection{\label{sec:level2} Distinguishing  two FM phases}
 In one-dimensional systems, the nearest-neighbor distance between magnetic atoms plays a crucial role in realizing fragile magnetism. According to the Stoner model for itinerant ferromagnet LaCrGe$_3$, pronounced peaks in the density of states are associated with the FM$_1$ and FM$_2$ phases \cite{nguyen2018using,niu2020evidence,wysokinski2019mechanism}, which can also be tuned via variations in thermal parameters. Temperature-dependent XRD measurements have also previously revealed anomalies in the lattice parameters near magnetic transitions, further supporting this behavior \cite{gati2021formation}. To obtain clearer evidence of the two FM phases, we performed temperature-dependent continuous Hall resistivity measurements at various applied magnetic fields in the FCW mode. In the paramagnetic state, $\rho_{yx}$ remains small because of the contribution only from normal Hall effect. Under an applied field of 0.01 T parallel to the \textit{z}-axis, $\rho_{yx}$ shows a steep increase near 85 K ($T_\mathrm{C}$) upon cooling, coinciding with the emergence of a spontaneous magnetization-induced anomalous Hall response. This is followed by a dip-like anomaly that correlates with a reduction in magnetization, due to magnetic domain depinning effect as discussed earlier.  Upon further cooling, $\rho_{yx}$ displays a small upturn before showing steady decrease down to the lowest measured temperature, characteristic of metallic transport behavior. The small upturn can be attributed to the rise in magnetization at low temperature. These features correspond to two distinct transitions, at $T_\mathrm{C}$ and $T_\mathrm{x}$, associated with the PM-FM$_1$ and FM$_1$-FM$_2$ transitions, respectively [Fig. 4(a)]. The transition temperatures shift to lower values due to the stabilization of a single-domain state as the magnetic field is increased from 0.01 T to 0.05 T and are completely suppressed above the saturation field (as seen in the curve at 5 T). The earlier magnetization measurements did not provide conclusive evidence for a spin-reorientation transition, and the presence of two ferromagnetic phases can instead be understood within the framework of domain-wall motion. Magneto-optical Kerr effect (MOKE) imaging reveals two distinct domain patterns below and above 70 K, corresponding to the two FM states \cite{ullah2023magnetic}. The difference in domain-wall widths between these regimes leads to two separate magnetically ordered states, each of which requires a different magnetic field to achieve full alignment. From the field-dependent resistivity $\rho_{xx}(B)$, it is difficult to identify the FM$_2$ transition, the data mainly shows a suppression of $T_\mathrm{C}$ under an applied magnetic field (see Figs. S4(e) and S4(f)). In contrast, the temperature-dependent magnetoresistance, defined as, ($\text{\textit{MR}}=(\rho_{xx}(B)-\rho_{xx}(0))/\rho_{xx}(0)$), measured for the applied fields of 1 T and 2 T along the \textit{y}-axis, reveals two distinct anomalies occurring nearly identical temperatures (Fig. 4(b)). This enhanced sensitivity of \textit{MR} compared to $\rho_{xx}(0)$ or $\rho_{xx}(B)$ indicates that the FM$_1$-FM$_2$ transition predominantly influences field-induced transport through modifications of domain-wall scattering or field-induced spin-dependent scattering. In general, every FM system shows pinned state at very low temperature which is characterized by the largest coercivity at the lowest temperature. However, LaCrGe$_3$ shows another pinned state at high temperature just below $T_\mathrm{C}$ corresponding to FM$_1$ phase. Between the two pinned states, the domain-wall depinning state enhances spin-spin interactions, thereby increasing resistivity, as evidenced in the  $\text{\textit{MR}}\%$. From Hall resistivity against magnetic field ($\rho_{yx}$-\textit{B} loop), we estimate coercive field ($B_\mathrm{c}$), and remanent Hall resistivity $\rho_{yx}|_{Sat}$, both of which peak around 73 K. In Figs. 4(c) and 4(d) FM$_1$-FM$_2$ transitions at $T_\mathrm{x}$ are shown from variation of $\rho_{yx}|_{Sat}$ and $B_\mathrm{c}$ as a function of temperature. For a $\rho_{yx}$-\textit{B} loop,  $\rho_{yx}|_{Sat}$ and $B_\mathrm{c}$ are indicated in Fig. 4(e). $R_0$ is calculated from the relation, ${\rho_{yx}/\textit{B} = R_0 + \mu_0R_SM/B}$, where $R_0$ is the intercept of plot $\rho_{yx}/\textit{B}$ vs ${M/B}$ in the high field region. For $B || y$, $R_0$ is nearly constant below 50 K, decreases sharply up to 73 K accompanied by a slope change from negative to positive, and then increases, exhibiting a slight slope change near the FM$_1$-PM transition ($\sim$85 K). Different slopes of $R_0$ across FM$_1$-FM$_2$ transition may occur due to the reconstruction of the Fermi surface near the transition temperature ($T_\mathrm{x}$) \cite{sandeman2003ferromagnetic,doiron2007quantum}. Hence, Hall effect proves to be a potent tool to distinguish two FM phases in LaCrGe$_3$. PM, FM$_1$ and FM$_2$ phases are separated by different colors in the Fig. 4(f). It is important to note that, for \textit{$B || z$}, the slope of $\rho_{yx}$, i.e, $R_0$ changes its sign to positive at higher temperature in the paramagnetic region while it remains negative over a broad temperature range for \textit{$B || y$} [in Fig. 2(a)].
\begin{figure}[h]
    \centering
    \includegraphics[width=0.5\textwidth]{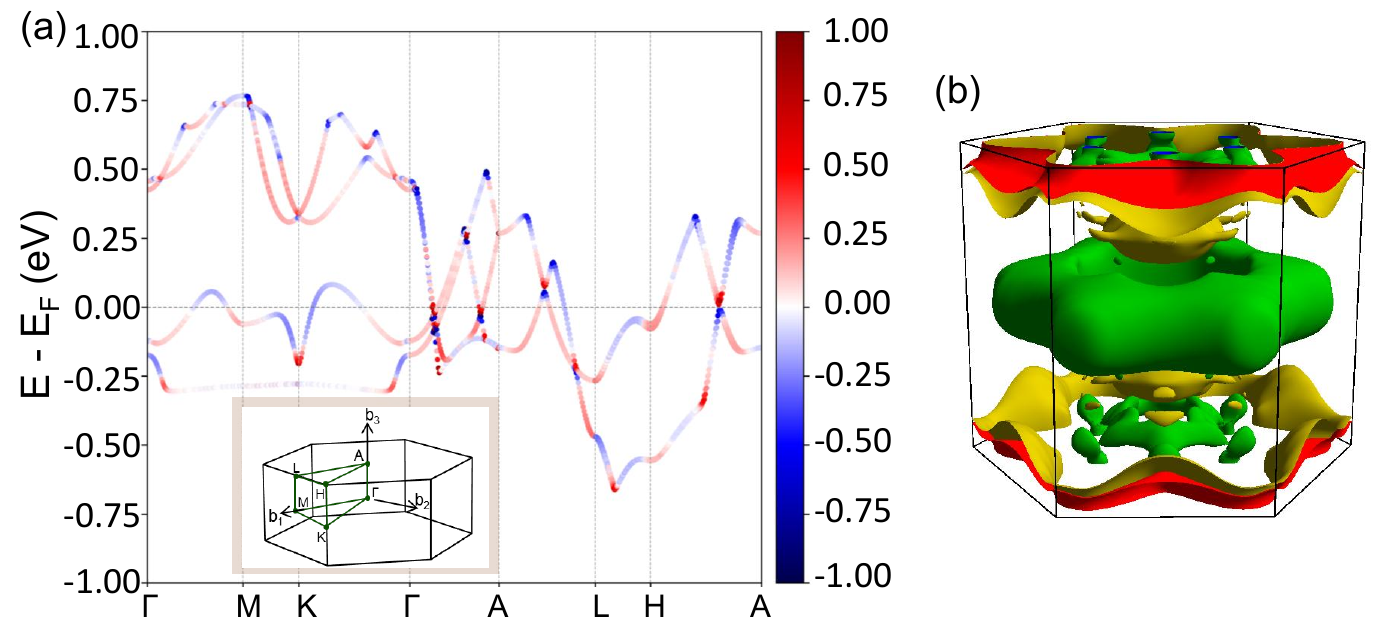}
    \caption{(a) Band structure plot of LaCrGe$_3$ with SOC, including only the bands that cross the Fermi energy (used to construct the Fermi surface), with a color scale representing the second-order derivative of the band energy with respect to crystal momentum. The maximum positive curvature is normalized to +1, and the maximum negative curvature is normalized to --1. (b) The Fermi surfaces corresponding to all four distinct bands are shown.}
    \label{Figure5}
\end{figure}
\subsection{\label{sec:level2}Band Structure Calculations}
{The crystal structure of LaCrGe$_3$ reveals that along the \textit{c}-axis a quasi-one-dimensional chain is formed by interconnected Cr-Ge octahedra, whereas in the \textit{ac} plane the alternating La-Ge and Cr layers produce a quasi-two-dimensional framework (Fig. 1(a)), and the anisotropic orbital hybridization along the different directions of the crystal leads to mixed structural dimensionality. Therefore, we expect a highly anisotropic electronic behavior due 
to mixed dimensional Fermi surface \cite{manako2024large}.}

For further understanding of the electronic structure, we computed the theoretical electronic band dispersion using the LDA exchange correlation functional, which shows better agreement with the experimental data. We have included SOC for all the data presented in this work. In Fig. \ref{Figure5}(a), we highlight only four bands that cross the \textit{E}$_{\text{F}}$ and the inset showing high symmetry points and paths in the first Brillouin zone. To determine the electron-like or hole-like character of the bands, we analyze the curvature (second derivative) of band energy with respect to crystal momentum.  
Color of the bands indicates the energy curvature at a given $\textit{k}$ and variation of the curvature (color gradient) from positive to negative indicates the character of the carrier transitioning from electron- to hole-like.

 \vspace{2 mm}Gaps open up for several band crossings near \textit{E}$_{\text{F}}$ in the presence of SOC. Bands cross the \textit{E}$_{\text{F}}$ multiple times along different high symmetry paths, resulting in a complex and multi-sheeted Fermi surface across the entire Brillouin zone as shown in Fig. \ref{Figure5}(b).  Assigned colors of the Fermi-surfaces correspond to differently colored bands as shown in Fig. S9(a) \cite{supplementary}. The band along the $\Gamma$-M-K-$\Gamma$ direction exhibits both positive and negative curvature in the vicinity of the Fermi energy \textit{E}$_{\text{F}}$, indicating mixed electron- and hole-like carrier behavior along this high-symmetry path, whereas, along the $\Gamma$-A, these bands exhibit predominantly positive curvature which means the carriers along this direction in these bands behave like electrons. In contrast, the bands crossing \textit{E}$_{\text{F}}$ along the A-L-H-A display mixed curvature, giving rise to mixed electron- and hole-like character. Character of the carriers along the longitudinal direction, which corresponds to $\Gamma$-A direction, is predominantly electrons. However, the three-dimensional, multi-sheeted Fermi surface exhibits point-to-point variations in local curvature and may give rise to both hole-like and electron-like behavior along this direction as well. In the \textit{ab}-plane which corresponds to $\Gamma$-M-K direction, carrier character is mixed. In summary, the detailed band structure and Fermi-surface geometry derived from DFT calculations indicate highly anisotropic carrier behavior, which is anticipated to manifest in the magnetotransport response.
\begin{figure}[h]
    \centering   \includegraphics[width=0.5\textwidth]{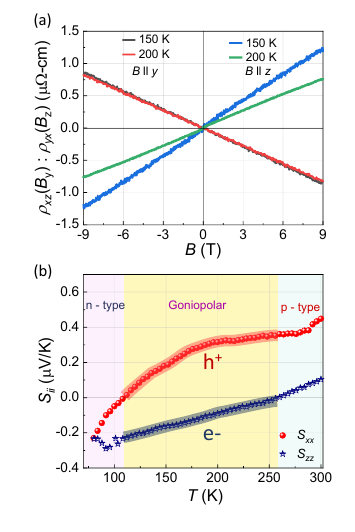}
    \caption{(a) Magnetic field dependent Hall resistivity in-plane ($\rho_{yx}(B_z)$) and out-of-plane ($\rho_{xz}(B_y)$) for the temperatures 150 K and 200 K. (b) Seebeck coefficient in-plane ($S_{xx}$) and out-of-plane ($S_{zz}$) for the temperatures 77 K to 300 K where the n-type, goniopolar and p-type regions are indicated by different colors.}
    \label{fig:your-label}
\end{figure}
\subsection{\label{sec:level2} Signatures of Goniopolarity in Seebeck and Hall effect}
The Seebeck and Hall effects are both highly sensitive to the geometry of the Fermi surface, as only the electronic states near the Fermi energy contribute significantly to transport properties. Depending on the nature of the Fermi surface, materials can show opposite charge carrier polarities along different crystallographic directions \cite{he2019fermi,wang2020chemical,yordanov2019large,nakamura2021axis,rai2025direction,helman2021ordinary,luo2023direction,koster2023axis,nelson2023axis,goto2024band,ochs2024synergizing,rowe1970thermopower,chung2004new,felser1998electronic,ochs2021computationally,manako2024large}. Such behavior can originate either from a single-band mechanism \cite{he2019fermi,wang2020chemical,yordanov2019large,nakamura2021axis}, known as goniopolarity, as observed in NaSn$_2$As$_2$ \cite{he2019fermi,nakamura2021axis} or from a multi-carrier mechanism \cite{rai2025direction,helman2021ordinary,luo2023direction,koster2023axis,nelson2023axis,goto2024band,ochs2024synergizing,rowe1970thermopower,chung2004new,felser1998electronic,ochs2021computationally,manako2024large}, as seen in CrSb \cite{rai2025direction}, where both electrons and holes from distinct bands dominate conduction along different axes. Fig. 6(a) presents the field-dependent $\rho_{ij}$ of hexagonal LaCrGe$_3$ at 150 K and 200 K for both in-plane and out-of-plane Hall geometry. When the magnetic field is applied along the out-of-plane direction (\textit{$B || z$}), the $\rho_{yx}(B_z)$ exhibits a positive slope, indicating dominant hole-type conduction. In contrast, for the in-plane direction (\textit{$B || y$}), the $\rho_{xz}(B_y)$ shows a negative slope, signifying dominant electron-type conduction. These observations clearly indicate the presence of direction-dependent charge carrier polarities, a hallmark of goniopolar transport \cite{he2019fermi}.

The Seebeck effect, a direct probe for determining the charge carrier type in a material, requires no application of the external magnetic field. Experimentally, the Seebeck coefficients $S_{ii} = -\,\frac{\Delta V_i}{\Delta T}$, quantify the longitudinal voltage developed along a given crystallographic direction in response to an applied temperature gradient along the same direction. The Seebeck coefficient ($S_{ii}$) is positive for hole dominant conduction and negative for electron dominant conduction. As shown in Fig. 6(b), the temperature-dependent Seebeck coefficients, measured along the in-plane ($S_{xx}$) and out-of-plane ($S_{zz}$) directions, exhibit contrasting behaviors \cite{behnia2015fundamentals}. Specifically, $S_{xx}$ is positive above 108 K, while $S_{zz}$ is negative below 257 K. Within this temperature range, $S_{xx}$ and $S_{zz}$ exhibit opposite signs, as shown in Fig. 5(b). The Seebeck coefficient can be expressed as \cite{he2019fermi}:
{\small
\begin{eqnarray}
S_{ii} = -\frac{\pi^2 k_B^2 T}{3 |e|} \left[ \frac{1}{n(E)} \frac{dn(E)}{dE} + \frac{1}{\tau_{ii}(E)} \frac{d\tau_{ii}}{dE} + \right.\nonumber \\ 
\left.m_{ii}^* \frac{d}{dE} \left( \frac{1}{m_{ii}^*}\right) \right]_{E=E_F}
\end{eqnarray}
}where \textit{n(E)} is the energy-dependent density of states and $\tau_{ii}(E)$ is the energy-dependent scattering time. The first two terms in Eq. (1) cannot lead to a sign reversal of $S_{ii}$ along different directions. Also the energy derivative of inverse effective mass $(m^{\ast}_{ii}\frac{d}{dE}(\frac{1}{m^{\ast}_{ii}}))$ is always negative, hence the anisotropic sign of S$_{ii}$ comes from the anisotropy in the sign of effective mass. The curvature of the electronic bands near the Fermi energy ($E_F$) determines the sign of $m_{ii}^*$. Bands with positive curvature (electron-like) yield a negative Seebeck coefficient, whereas those with negative curvature (hole-like) result in a positive Seebeck coefficient \cite{he2019fermi}. In systems with both concave and convex Fermi surface regions, such as those with hyperboloid-shaped Fermi surfaces, the sign of $m_{ii}^*$ can vary with direction, resulting in the observed anisotropic thermoelectric behavior \cite{jan1968effective,behnia2015fundamentals}.

 The theoretically calculated highly anisotropic and intricate Fermi surface (in Fig. 5(b)) geometry supports the coexistence of both electron and hole type carriers. Although, the dominant contribution to electrical conduction for the particular direction also depends on carrier mobilities. A detailed information on the Fermi surface geometry is provided in \cite{supplementary}. Taken together, the Hall and Seebeck effect along with DFT calculations, we can conclude  LaCrGe$_3$ as a goniopolar material, characterized by the conduction polarity that reverses depending on the crystallographic direction. This highlights LaCrGe$_3$ as a promising candidate for as novel thermoelectric material for practical device applications \cite{scudder2021highly,tang2015p}.
 
\section{SUMMARY}
In this work, we have performed a detailed magneto-transport properties of the tri-critical wing compound LaCrGe$_3$. Hall effect measurements discern two ferromagnetic phases clearly in this compound. In the presence of small magnetic fields below the saturation field, Hall resistivity shows two distinct transitions, for PM-FM$_1$ and FM$_1$-FM$_2$. In addition, the temperature dependence of $B_\mathrm{c}$, $\theta_\mathrm{AH}$, $\rho_{yx}|_{Sat}$, and $R_0$ provides significant changes near the transition between two ferromagnetic phases. This suggests that the electronic band structure is sensitive to such phase changes that can be detected by transport measurements.  We have extended the Hall effect study in the paramagnetic region which uncovers a goniopolarity or direction-dependent conduction polarity in this compound. For a large paramagnetic temperature range, the type of majority carrier depends on the direction of the applied magnetic field, which is further supported by the Seebeck effect wherein we find opposite signs for the Seebeck coefficient for the in-plane and out-of-plane applied temperature gradient. First principles calculations reveal a large anisotropy in the Fermi surface which supports the observation of  the change of conduction polarity. Goniopolarity and complex magnetism coupled with peculiar domain wall dynamics make LaCrGe$_3$ an important candidate for advances in electronics devices.
\begin{acknowledgments}
This research was carried out using the instrumentation facilities at the Technical Research Centre (TRC), S. N. Bose National Centre for Basic Sciences, supported by the Department of Science and Technology (DST), Government of India. N.K. acknowledges financial support from the Science and Engineering Research Board (SERB), India, under Grant No. CRG/2021/002747, as well as funding from the Max Planck Society through the Max Planck-India Partner Group program. S.S. acknowledges financial support from ANRF, SERB, India, Grant No. CRG/2023/002082. M.S. acknowledges the DST, India, for support through a fellowship. N.I. acknowledges the University Grant Commission (UGC), India, for support through a fellowship. A.J. and M.K. acknowledge National Supercomputing Mission (NSM) for providing computing resources of ‘PARAM RUDRA’ at S. N. Bose National Centre for Basic Sciences, which is implemented by C-DAC and supported by the Ministry
of Electronics and Information Technology (MeitY) and Department of Science and Technology (DST), Government of India.
\end{acknowledgments}
\section*{Data Availability}
The data supporting the findings of this study cannot be made publicly available at the time of publication due to technical limitations and the prohibitive cost of preparing, depositing, and hosting the datasets. However, the data can be obtained from the authors upon reasonable request.


\providecommand{\noopsort}[1]{}\providecommand{\singleletter}[1]{#1}%

\end{document}